# Pressure induced nonmonotonic evolution of superconductivity in 6$R$-TaS$_2$ with natural bulk van der Waals heterostructure

Shaopeng Wang,[1,2] Yuyan Han,[1] Sutao Sun,[3] Shuyang Wang,[1] Chao An,[4] Chunhua Chen,[1] Lili Zhang,[5] Yonghui Zhou,[1,2,a)] Jian Zhou,[3,a)] and Zhaorong Yang[1,2,4,6,a)]

[1]*Anhui Key Laboratory of Low-energy Quantum Materials and Devices, High Magnetic Field Laboratory, HFIPS, Chinese Academy of Sciences, Hefei 230031, China*

[2]*Science Island Branch of Graduate School, University of Science and Technology of China, Hefei 230026, China*

[3]*National Laboratory of Solid State Microstructures and Department of Materials Science and Engineering, Nanjing University, Nanjing 210093, China*

[4]*Institutes of Physical Science and Information Technology, Anhui University, Hefei 230601, China*

[5]*Shanghai Synchrotron Radiation Facility, Shanghai Advanced Research Institute, Chinese Academy of Sciences, Shanghai 201204, China*

[6]*Collaborative Innovation Center of Advanced Microstructures, Nanjing 210093, China*

[a)]Corresponding authors: yhzhou@hmfl.ac.cn, zhoujian@nju.edu.cn and zryang@issp.ac.cn

**Abstract**

The natural bulk van der Waals heterostructures compound 6$R$-TaS$_2$ consists of alternate stacking 1$T$- and 1$H$-TaS$_2$ monolayers, creating a unique system that incorporates charge-density-wave (CDW) order and superconductivity (SC) in distinct monolayers. Here, after confirming that the 2D nature of the lattice is preserved up to 8 GPa in 6$R$-TaS$_2$, we documented an unusual evolution of CDW and SC by conducting high-pressure electronic transport measurements. Upon compression, we observe a gradual suppression of CDW within the 1$T$-layers, while the SC exhibits a dome-shaped behavior that terminates at a critical pressure $P_c$ around 2.9 GPa. By taking account of the fact that the substantial suppression of SC is concomitant with the complete collapse of CDW order at $P_c$, we argue that the 6$R$-TaS$_2$ behaves like a stack of Josephson junctions and thus the suppressed superconductivity can be attributed to the weakening of Josephson coupling associated with the presence of CDW fluctuations in the 1$T$-layers. Furthermore, the SC reversely enhances above $P_c$, implying the development of emergent superconductivity in the 1$T$-layers after the melting of $T$-layer CDW orders. These results show that the 6$R$-TaS$_2$ not only provides a promising platform to explore emergent phenomena but also serves as a model system to study the complex interactions between competing electronic states.

The advancements in the isolation and manipulation of atomically thin sheets of two-dimensional (2D) materials have paved the way for a new era of fundamental scientific research and technological innovation [1-4]. Nowadays, it has become feasible to independently fabricate distinct single layers possessing specific properties and subsequently assemble them into novel quantum materials known as van der Waals heterostructures (vdWHs). The constructed vdWHs not only combine the respective materials' functionalities but also imprint exotic properties through proximity interactions across the interface, leading to emergent phenomena such as Hofstadter butterfly states, tunable Mott insulators and topological superconductivity, etc [5-11].

Recently, there has been a growing interest in the exploration of emerging natural bulk vdWHs among transition-metal dichalcogenides [12-18]. Similar to the building blocks in 2D vdWHs, these bulk vdWHs are built as alternate stacking of octahedral (*T*) layers with charge-density-wave (CDW) order and trigonal-prismatic (*H*) layers with superconductivity (SC) and/or different CDW orders. To date, the bulk vdWHs are mainly concentrated on 4$H_b$- and 6$R$-TaX$_2$ (X = S, Se) [12-18]. For the 4$H_b$ phase, taking 4$H_b$-TaS$_2$ as an example, the unit cell consists of four alternate stacking monolayers of 1$T$-TaS$_2$ and 1$H$-TaS$_2$ (half of 2$H$-TaS$_2$) [13]. The bulk 1$T$-TaS$_2$ is a Mott insulator [19] and candidate gapless quantum spin liquid [20], while the bulk 2$H$-TaS$_2$ is a superconductor known to also host a competing CDW order [21]. The natural combination of these two strongly correlated systems with proximity effect can generate fascinating physical phenomena like chiral SC, topological surface SC, and spontaneous vortices with a hidden magnetic phase [13, 16, 17]. As for the 6$R$ phase, the typical compound 6$R$-TaS$_2$ is composed of six alternate stacking 1$H$- and 1$T$-TaS$_2$ layers in a unit cell, see Fig. 1(a). Recently, it has been reported that 6$R$-TaS$_2$ undergoes two CDW transitions from the

$T$-layer (nearly commensurate CDW transition at 320 K and commensurate CDW transition at 305 K) and enters the superconducting state from the adjacent $H$-layer below 2.7 K at ambient pressure [12]. The natural incorporation of CDW and SC in distinct layers makes $6R$-TaS$_2$ stand out from the CDW materials, providing a unique platform to investigate the intriguing interplay between CDW and SC.

Pressure is proven to be an effective way of tuning the interlayer coupling and regulating the electronic properties in both artificial few-layer vdWHs and bulk TMDs [21-29]. In this work, we have investigated the pressure-induced evolution of the structural and electronic properties of $6R$-TaS$_2$ single crystals, and unveil an unusual interplay between CDW and SC therein. With increasing pressure, the CDW order is destabilized monotonically, while the SC displays a dome-like behavior followed by a subsequent enhancement at higher pressures. Specifically, along with the complete collapse of CDW order at a critical pressure of around 2.9 GPa, the SC undergoes a simultaneous suppression. Since no structural phase transition or dimension crossover occurs throughout the pressure enhancement, we argue that the unusual evolution of SC can be well accounted for in a scenario of pressure tuning of Josephson coupling.

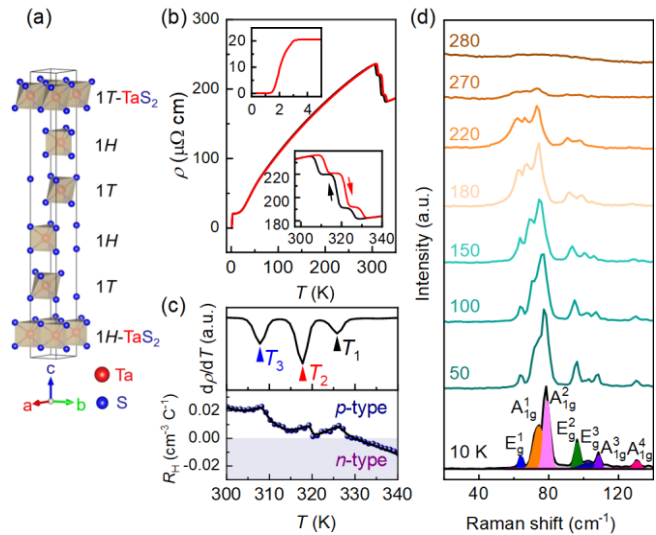

FIG. 1. (a) Schematic crystal structure of 6$R$-TaS$_2$. (b) Temperature-dependent resistivity curves at ambient pressure. Upper left inset: enlarged view around $T_c^{onset}$ ~ 2.8 K ($T_c^{onset}$ is determined by 90% of normal state resistivity). Lower right inset: thermal hysteresis across step-like resistivity upturns. (c) Temperature-dependent d$\rho$/d$T$ and Hall coefficient $R_H$ between 300 and 340 K. The characteristic temperatures of three step-like upturns, are denoted by $T_1$, $T_2$, and $T_3$, respectively. (d) Selected ambient-pressure Raman spectra from 10 to 280 K.

The synthesis and experimental details are given in the Supplemental Material [30]. Figure 1(b) shows the temperature-dependent resistivity $\rho(T)$ curves of 6$R$-TaS$_2$, which document a superconducting transition in the 1$H$-layer with $T_c^{onset}$ ~ 2.8 K, and step-like upturns in a temperature range of about 30 K associated with the CDW transitions in the 1$T$-layer, consistent with the results of Achari *et al.* [12]. The step-like anomalies display clear hysteresis upon cooling and warming [see the inset of Fig. 1(b)]. As seen in Fig. 1(c), upon cooling across the CDW transitions, the anomalies manifest as dips in the corresponding derivative d$\rho$/d$T$ at 326 ($T_1$), 318 ($T_2$), and 308 K ($T_3$), while the Hall coefficient $R_H$ changes sign from negative to positive followed by three peak-like anomalies at the corresponding temperatures. Note that Achari *et al.* have identified two CDW transitions in 6$R$-TaS$_2$ at 320 and 305 K, respectively, based on their electrical transport measurements [12]. The former was attributed to a transition from incommensurate to nearly commensurate (NC) CDW phase, and the latter was assigned to a NC- to commensurate (C) CDW transition [12]. Clearly, by alternately stacking the 1$T$- and 1$H$-layers, the low-temperature CCDW transition of the original 1$T$-layer is increased from 190 K in bulk to above room temperature in 6$R$-TaS$_2$ [25]. Meantime, the superconducting temperature in the 1$H$-layer is enhanced substantially compared to the value of 0.8 K in its bulk counterpart [43].

In addition to electrical transport measurements, the CDW state was also detected by Raman spectroscopy. Previous studies on the 1$T$-TaS$_2$ indicated that upon the CCDW transition, the zone-folding Raman modes manifest as a serial of A$_{1g}$ and E$_g$ Raman modes below 140 cm$^{-1}$, which are the signature of the CCDW state [44, 45]. Figure 1(d) displays the temperature-dependent Raman spectra at ambient pressure and selected temperatures from 10 to 280 K. At 10 K, the zone-folding Raman modes exhibit four A$_{1g}$ and three E$_g$ modes. All the zone-folding modes are softened as temperature increases, and the CCDW feature can be observed until 270 K.

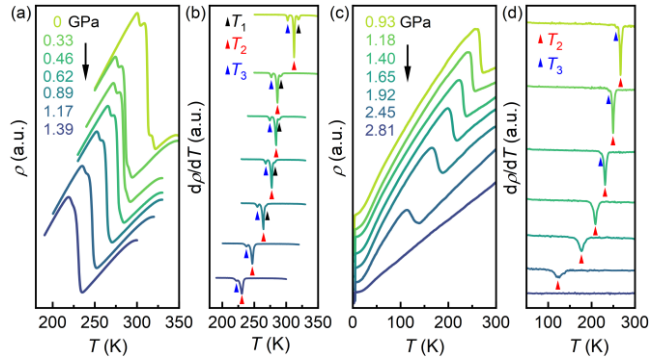

FIG. 2. Temperature-dependent resistivity of 6$R$-TaS$_2$ and its derivative d$\rho$/d$T$ in PCC [(a)-(b)] and DAC [(c)-(d)] setups. All the curves are offset for clarity (see the original data in Fig. S2).

After having characterized the sample quality, we first performed electronic measurements in a piston cylinder cell (PCC) setup, with Daphne 7373 as the pressure-transmitting medium (PTM) to examine the pressure-induced evolution of CDW. Figure 2(a) and 2(b) display the resistivity versus temperature $\rho(T)$ curves in the temperature range of 180 to 350 K and their derivative d$\rho$/d$T$ versus $T$ within 1.39 GPa. At the starting pressure of 0 GPa, we observed three distinct step-like resistivity upturns which are consistent with the ambient pressure case shown in Fig.

1. During compression, the resistive upturn is progressively shifted to lower temperatures, indicating the suppression of CDW order. All three characteristic temperatures evolve independently without converging except that the amplitude of the resistivity anomaly around $T_1$ is suppressed more dramatically and cannot be discerned at 1.17 GPa.

To get full insight into the pressure effect on CDW orders and unveil the relationship between CDWs and superconductivity, we further conducted high-pressure electrical transport measurements with the PTM Daphne 7373 in a diamond anvil cell (DAC) setup. As shown in Fig. 2(c), at the starting pressure of 0.93 GPa, the resistivity displays an abrupt upturn at $T_2$ associated with the CDW transitions. The resistivity anomaly at $T_3$ can be discerned only in the d$\rho$/d$T$ curve. With increasing pressure, the resistivity upturn moves to lower temperatures, in agreement with the results in Fig. 2(a). At 2.81 GPa, the step-like feature suddenly disappears and cannot be detected in the d$\rho$/d$T$ curve either, which suggests the complete suppression of CDW orders. The suppression of the CDW around 2.81 GPa is further confirmed by the evolution of high-pressure Raman spectra at 80 K (see Fig. S5). It is noteworthy that 80 K is the lowest temperature explored in the measurement due to the strong signal from the background of Daphne 7373. As shown in Fig. S5, along with the suppression of the CDW, all the zone-folding modes continuously soften and are undetectable above 2.71 GPa.

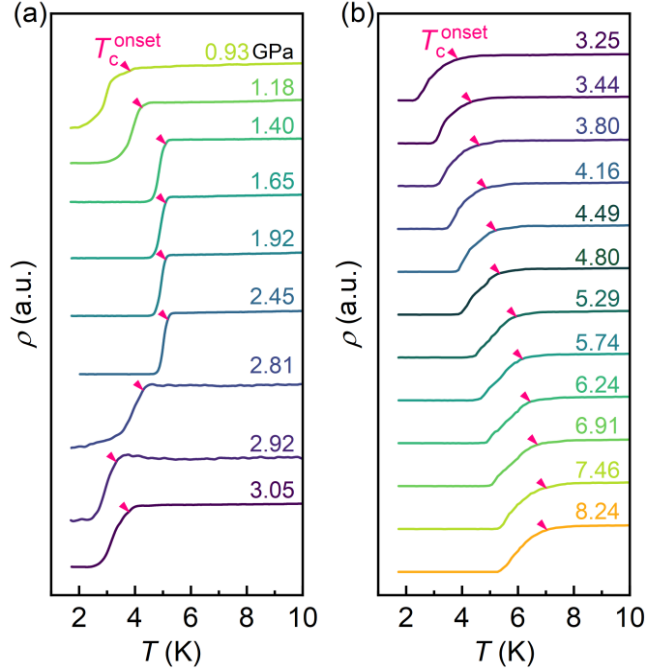

FIG. 3. (a)-(b) Low-temperature $\rho(T)$ curves in DAC setup. Pink solid triangle denotes the $T_c^{onset}$ at each pressure. The curves are offset for clarity.

Figure 3 displays the pressure-induced evolution of the SC, illustrating a nonmonotonic variation of the superconducting transition temperature versus pressure. Compared with ambient pressure, the superconductivity gets a pronounced promotion at 0.93 GPa with $T_c^{onset}$ = 3.79 K. The $T_c^{onset}$ continuously increases with pressure, attaining approximately 5.16 K at 1.65 GPa. After that, the $T_c^{onset}$ remains nearly constant. Interestingly, when reaching the pressure of CDW disappearance at 2.81 GPa, the $T_c^{onset}$ decreases remarkably, accompanied by pronounced broadening of superconducting transition. Furthermore, at 3.05 GPa and above, the superconductivity gets another unexpected promotion. The $T_c^{onset}$ keeps increasing and reaches the maximum values of 6.91 K at 8.24 GPa, the highest pressure in our transport measurement.

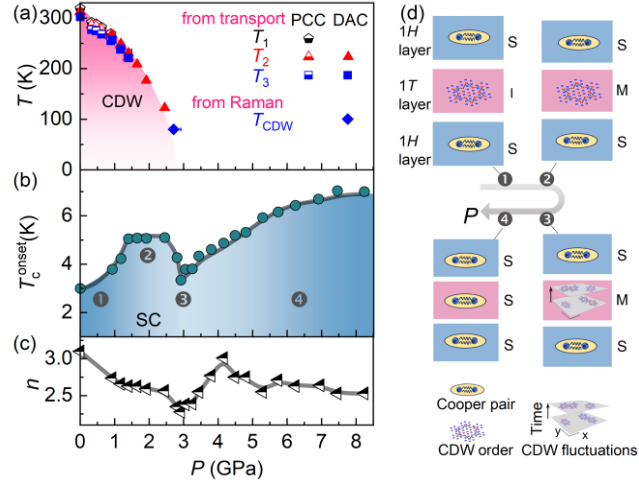

FIG. 4. Temperature-pressure phase diagram of 6$R$-TaS$_2$. Pressure dependence of (a) CDW transition temperatures, (b) superconducting $T_c^{onset}$, and (c) parameter *n* from the analysis of the normal state resistivity. (d) Schematic illustration of pressure-tuned Josephson coupling in vdW 6$R$-TaS$_2$ along with successive transformation of the 1$T$-layer from insulator (I) to metal (M) to metal with CDW fluctuations to superconductor.

Based on the above comprehensive high-pressure measurements, we constructed the temperature-pressure phase diagram of 6$R$-TaS$_2$ as depicted in Fig. 4, which directly illustrates the evolutions of CDW and SC as well as their correlations. As can be seen, the $T_{CDW}$ decreases monotonically to lower temperatures and extrapolates to zero at a critical pressure of around 2.9 GPa. Meantime, separated exactly by the same pressure, the $T_c^{onset}$-$P$ phase diagram reveals two distinct SC regions. The superconducting $T_c^{onset}$ initially increases to 5.16 K upon compression up to 1.65 GPa then levels off followed by a sharp depression above 2.45 GPa, forming a dome-shaped superconducting state terminating exactly at 2.9 GPa. Above 2.9 GPa, the superconductivity experiences a resurgence, with $T_c^{onset}$ eventually reaching the maximum of 6.91 K at 8.24 GPa. The presence of dome-shaped SC in response to external parameters like pressure is universal in the CDW materials. Associated with the competition between CDW and SC, the superconducting dome normally develops right after the suppression of the CDW or displays a maximum at the critical pressure when the CDW collapses completely [28, 46, 47]. In some other cases, the maximum $T_c$ is located in the vicinity of the quantum critical point, where the CDW fluctuations are responsible for the enhancement of SC [48-50]. In contrast, the dome in $6R$-TaS$_2$ ends up at the critical pressure corresponding to the disappearance of the CDW, which signals that the collapse of CDW is responsible for the synchronous suppression of SC.

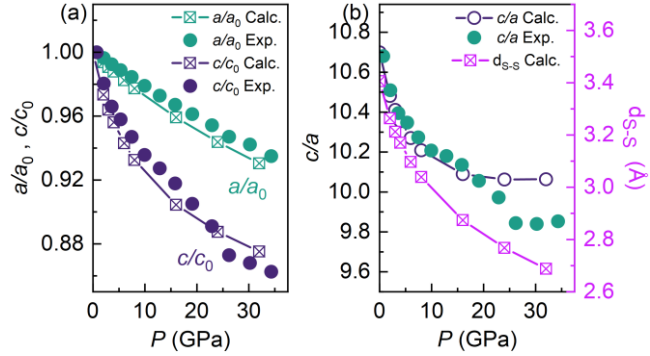

FIG. 5. (a) Experimental and calculated $a/a_0$, $c/c_0$ ratios and (b) experimental and calculated $c/a$ ratios and the calculated interlayer S-S distance (right axis) as a function of pressure. The $a_0$ and $c_0$ are the lattice constants of $6R$-TaS$_2$ at ambient pressure.

To clarify the unusual $T_c$ evolution under pressure, we performed *in situ* high-pressure synchrotron XRD measurements on $6R$-TaS$_2$. As shown in Fig. S6, no structural phase transition is observed under compression up to 34.3 GPa. Both the lattice parameters $c$ and $a$ decrease monotonically, while $c$ is compressed more pronounced because of the 2D nature of the lattice, which is also reflected in the pressure-dependent axial ratio $c/a$, see Fig. 5. Upon further compression above 25 GPa, $c/a$ ratio becomes nearly constant, implying that the interlayer interactions become rather strong. Similar behavior in the $c/a$ ratio has also been observed in pressurized CsV$_3$Sb$_5$ [51] and $4H_b$-TaSe$_2$ [14], and was attributed to the dimension crossover from the quasi-2D to the 3D structure under pressure.

To gain more insight into the evolution of the crystal structure under high pressure, we performed the first-principles calculations on the high-pressure crystal structure of 6$R$-TaS$_2$. The structure information at various pressures is given in Table S1. The calculated $a/a_0$, $c/c_0$, and $c/a$ are also plotted in Fig. 5(a) and 5(b). The evolution of the calculated parameters is well consistent with the experimental results below 8 GPa, but the $c/c_0$ and $c/a$ deviate significantly from the experimental values above 20 GPa, possibly due to the nonhydrostatic pressure effect at higher pressure. Meantime, the $c/a$ ratio rapidly decreases at low pressure, and the decay rate of $c/a$ exhibits a crossover above 8 GPa. After applying pressure, sulfur atoms between adjacent $H$- and $T$-layers get close to each other along the $c$-axis. The distance $d_{S-S}$ between the interlayer sulfur atoms is compressed from 3.55 Å at 0 GPa to 3.04 Å at 8 GPa and 2.87 Å at 16 GPa. Early work showed that the length of S-S bonds typically ranges between 1.8 and 3.0 Å [52], suggesting that below 8 GPa, 6$R$-TaS$_2$ maintains a vdW structure.

Without involving a structural transition, we will see that the unusual evolution of SC in pressurized 6$R$-TaS$_2$ is indeed rooted in the specific vdW 2D nature, which persists throughout the phase diagram in Fig. 4. At ambient pressure, the bulk 6$R$-TaS$_2$ is composed by alternate stacking of 1$H$- and 1$T$-TaS$_2$ layers, *i.e.*, alternating SC and CDW monolayers [12]. The superconductivity is not only determined by the paring within the 1$H$-layer, but also more relied on the interlayer coupling that gives rise to a global phase coherence. This scenario is very similar to that in the high-$T_c$ cuprate superconductors like Bi-2212, where the superconducting double layers are coupled by the Josephson effect, and the 6$R$-TaS$_2$ crystal is thus equivalent to a stack of superconductor-insulator-superconductor Josephson junctions [53, 54]. Taking into account the fact that the superconducting $T_c$ in 2$H$-TaS$_2$, *i.e.*, the bulk counterpart of 1$H$-layer, keeps continuous increasing below 8 GPa [28, 29, 55], we suggest that the pressure-induced nonmonotonic evolution of SC in 6$R$-TaS$_2$ is mainly dominated by the unusual variation of the Josephson coupling as schematically illustrated in Fig. 4(d).

In the initial stage, as shown in panel 1 of Fig. 4(d), the Josephson coupling is mediated by tunneling of the Cooper pairs through the insulating barrier [53, 54, 56]. The increase of applied pressure leads to the decrease of the interlayer distance on one hand, and results in the suppression of CDW or insulating behavior in the middle 1$T$-layer on the other hand. Both are favorable to the enhancement of Josephson coupling and thereby improve the superconductivity.

The enhancing trend of superconductivity stops if the middle 1$T$-layer becomes metallic with increasing pressure [panel 2 in Fig. 4(d)]. In the case of SC-metal-SC junction, the Josephson coupling is accordingly established by the superconducting proximity effect rather than by the tunneling of Cooper pairs [56-61]. The proximity effect is characterized by inducing a superconducting order parameter in the metallic layer and weakening the superconductivity within a distance of the order of the coherence length in the SC layer [57, 59, 60]. The transformation of the coupling way by the proximity effect could be the main reason for the presence of the plateau between 1.65 and 2.45 GPa, otherwise, the $T_c^{onset}$ should increase monotonically. In support of this speculation, as shown in Fig. S3, the resistivity along $c$-axis at 10 K drops by about an order of magnitude with increasing pressure from 0 to 1.62 GPa, above which it tends to be stable against the pressure.

From a microscopic point of view, the superconducting proximity effect is correlated to the Andreev reflection, *i.e.*, reflecting an incident electron as a hole in the metallic layer and transmitting a Cooper pair into the SC region correspondingly [56, 59, 60, 62]. Around 2.9 GPa, along with the disappearance of the CDW order in the 1$T$-layer, the superconductivity displays a synchronous depression. Furthermore, the exponential parameter $n$, extracted from the fitting of $\rho(T)$ curves in the normal state with the empirical formula $\rho(T) = \rho_0 + AT^n$, also exhibits a dip feature around this critical pressure, see Fig. 4(c). The suppression of the parameter $n$ can be viewed to be evidence of strong CDW fluctuations accompanied by the collapse of CDW order [21, 24, 48]. In the presence of CDW fluctuations, as illustrated in panel 3 of Fig. 4(d), it is plausible to expect that the transparency of the junction becomes lower because the CDW fluctuations give rise to critical scattering for the incident electron and hamper the realization of Andreev reflection, leading to the suppression of proximity effect [60].

Further increase of pressure beyond the first SC dome resulted in a subsequent resurgence of superconductivity, which can be attributed to the development of emergent superconductivity in the 1$T$-layers by referencing the studies of pressure-induced superconductivity in the bulk 1$T$-TaS$_2$ [24, 25]. Along with the presence of stronger superconductivity in the 1$T$-layers [see panel 4 of Fig. 4(d)], the scenario of stacking Josephson junctions is not applicable anymore, which should involve a dimension enhancement of the superconductivity.

In summary, by compressing the vdWHs compound that maintains its 2D vdW nature through the pressure enhancement in the $T$-$P$ phase diagram, we documented an unusual interplay between SC and CDW. Associated with the modulation of inherent Josephson coupling under compression, the SC undergoes successive variations: initial increases, plateaus between 1.65 and 2.45 GPa, depresses around 2.9 GPa, and finally resurges at higher pressures. The Josephson coupling is highly suppressed at 2.9 GPa because of the presence of metallic conductivity and CDW fluctuations in the 1$T$-layers. Our findings of the CDW-fluctuation-induced suppressing of SC in

$6R$-TaS$_2$ may shed light on studying the interplay between competing electronic orders in the vdWHs.

## ACKNOWLEDGMENTS

This work was supported by the National Key Research and Development Program of China (Grants No. 2023YFA1406102 and 2022YFA1602603), the National Natural Science Foundation of China (NSFC) (Grants No. 12374049, No. 12174395, No. 12204004, and No. 12004004), the Natural Science Foundation of Anhui Province (Grants No. 2308085MA16 and No. 2308085QA18). Yonghui Zhou was supported by the Youth Innovation Promotion Association CAS (Grant No. 2020443). A portion of this work was supported by the Basic Research Program of the Chinese Academy of Sciences Based on Major Scientific Infrastructures (No. JZHKYPT-2021-08). A portion of this work was performed on the Steady High Magnetic Field Facilities, High Magnetic Field Laboratory, CAS. The numerical calculations in this paper have been done on the computing facilities in the High Performance Computing Center (HPCC) of Nanjing University.